# Mass Distribution and "Mass Gap" of Compact Stellar Remnants in Binary Systems

Niranjan Kumar[1*] and Vladimir V. Sokolov[2]

[1] *Institute for Physics of Microstructures, Russian Academy of Sciences, Nizhny Novgorod, 603087 Russia.*
[2] *Special Astrophysical Observatory, Russian Academy of Sciences, Nizhnii Arkhyz, 369167 Russia.*

**Abstract** – The highest critical mass of neutron stars (NSs) was reviewed in the context of equation of state and observational results. It was predicted that the maximum NS mass ($M_{NS}$) exists in the range $M_{NS} \approx 2.2$–$2.9\ M_\odot$. However, recent observations of gravitational waves and other studies had suggested the higher mass limit of NSs, $M_{NS} \approx 3.2\ M_\odot$. The NS mass up to the value of $M_{NS} \approx 2\ M_\odot$ is well understood, and with such a mass value it was meaningful to discuss the "mass gap" ("m-gap") between the NS and black hole (BH) collapsars. The "m-gap" exists in between the highest mass of NS and the lowest mass of BH collapsars ($M_{\text{m-gap}} \approx 2$–$5\ M_\odot$). In the mass distribution, the maximum population of NSs and BHs is located at $M_{NS} = 1.4\ M_\odot$ and $M_{BH} = 6.7\ M_\odot$, respectively. However, recent observational results predicted filling the "m-gap" by the compact objects. In this paper, the concept of gravidynamics was reported to resolve the problem of peak likelihood value of gravitational mass at $M_{\text{peak}} = 6.7\ M_\odot$ and the "m-gap" ($M_{\text{m-gap}} \approx 2$–$5\ M_\odot$). This concept was based on a non-metric scalar-tensor model of gravitational interaction with localizable field energy. The gravidynamics model shows the total mass ($M_Q$) of a compact relativistic object filled with matter of quark-gluon plasma of the radius $r^* = GM_Q/c^2 \approx 10$ km, consistent with the "m-gap". It was conceptualized that the total measurable gravitational mass of such an extremely dense object consists of both matter and field, which is described by scalar-tensor components. This model is also useful for predicting the collapsars within the "m-gap".

**Key words:** black hole physics – gravitational waves – binary stars: compact – stars: black holes – stars: neutron – stars: collapsars – stars: stellar mass distribution – gravitation theory: gravidynamics

E-mail: kumar@ipmras.ru

## 1. INTRODUCTION

The compact stellar remnants in close binary systems (CBSs) have been extensively studied (Abbott et al. 2019, 2020a,b,c, 2021a; Belczynski et al. 2012; Farr et al. 2011; Fragos et al. 2008, 2009; Irwin 2006; Kiziltan et al. 2013; Kreidberg et al. 2012; Littenberg et al. 2015; Mandel et al. 2015; Margalit and Metzger 2017; Özel et al. 2010, 2012; Petrov et al. 2014; Sokolov 1992, 2015, 2019), and the physical properties of such astrophysical objects are still the subject of research. These properties include type of matter and phase of the compact dense object, thermodynamic properties and the exact value of equation of states (EoSs). In the series of problems, the most puzzling question about CBSs is related to the mass distribution and the "mass gap" (m-gap) (Abbott et al. 2020a,c, 2021b; Bailyn et al. 1998; Kreidberg et al. 2012; Özel et al. 2010, 2012; Petrov et al. 2014; Wyrzykowski and Mandel 2020; Yang et al. 2020). Abbott et al. (2021b) have reported in the third Gravitational-wave Transient Catalog (GWTC-3) the range of inferred component masses of compact objects similar to that found with previous catalog (GWTC-2.1). The O3b candidates include the first confident observations of neutron star (NS) – black hole (BH) binaries. However, it was pointed out that from the gravitational-wave data alone, the measurement of matter effects is not possible that distinguishes whether the binary components are NSs or BHs. The properties of gravitational wave data are almost similar in case for the highest and lowest mass of NS and BH, respectively.

The implications of these issues for astrophysical observations may answer deeper questions than model-based studies. These questions are mainly related to the core collapse supernova (CCSN) explosion mechanism (Burrows and Vartanyan 2021), the composition of particles, and the energy of the gravitational field in the NS core, which can also help to accurately obtain the EoSs. The CCSN can also be useful for understanding the gravitational potential and predicting the space-time geometry of NSs and BHs.

A number of observational and theoretical models showed the considerable "m-gap" in the mass distribution existing between the highest mass of NS and lowest mass of BH (Abbott et al. 2020a,c; Belczynski et al. 2012; Kreidberg et al. 2012; Littenberg et al.2015; Mandel et al.2015; Özel et al.2010, 2012; Wyrzykowski and Mandel 2020; Yang et al. 2020; Zevin et al. 2020). Such an observed "m-gap" between the NSs and BHs exists in the mass range of $M_{\text{m-gap}} = 2–5\ M_\odot$. It seems that the gravitational mass of NSs and BHs follows the explicit mass values. The properties of such compact collapsars depend on the mass, density, and EoS. Specific values of the mass are clearly visible in 16 CBSs with low-mass optical companions with a mass of $0.6\ M_\odot$ (Kreidberg et al. 2012; Özel et al. 2010, 2012). In these binaries, the two separate peaks related to the masses of NSs and BHs in the mass distribution was shown. It was found that a peak in the mass distribution of relativistic objects is close to $M_{\text{peak}} = 6.7\ M_\odot$ (Kreidberg et al. 2012; Özel et al. 2010, 2012). However, recent studies showed that the "m-gap" in the distribution is filled by compact dense objects, which are claimed to be close in properties to NSs or BHs (Abbott et al. 2020a,c; Belczynski et al. 2012; Fryer et al. 2002; Fryer and Kalogera 2001; Gelino and Harrison 2003; Gupta et al. 2020; Thompson et al. 2019). Such observations are disputed, and a firm explanation of our understanding of the properties of such matter, existing in the "m-gap" region, has been not sufficiently well studied. The distribution of masses in CBSs, specific values of mass of compact objects, existence of "m-gap" and peak of relativistic object at $M_{\text{peak}} = 6.7\ M_\odot$ can be better understood by finding the relationship between CCSN explosions and gamma-ray bursts (GRBs). This relationship can also provide information about the CCSN explosion mechanism itself (Burrows and Vartanyan 2021). Polarized radiation of long GRBs, the black-body component in their spectra and other essential observational properties can be explained also by a direct manifestation of collapsars (NSs and BHs).

In this article, we discuss the mass distribution of NSs and BHs in compact binary systems and find out the possible cause of the "m-gap" between NSs and BHs. The mass distribution of CBSs was presented taking into account the binary systems NS-NS, NS-BH, NS-white dwarf (WD) and NS-companions of low-mass binary stars. The existence of dense compact objects filling the "m-gap" region is also proposed and discussed. A closer examination of the peak value of the gravitational mass at $M_{peak} = 6.7\ M_\odot$ in the observed mass distribution of collapsars is emphasized. A plausible interpretation of such a peak value of the gravitational mass is proposed within the framework of a direct observational consequence of localizable gravitational field energy in scalar-tensor gravidynamics.

## 2. PHYSICAL LIMITATIONS OF THE MASS OF NEUTRON STARS

This section is devoted to understanding the large masses of NSs, which is considered as the most important criteria for describing the basis of the "m-gap" problem.

Since the discovery of neutrons, the mass distribution of NSs has been at the center of attention of compact object for astrophysicist Baade and Zwicky (1934a,b,c); Chandrasekhar (1931); Landau (1932); Shapiro and Teukolsky (1983). The works of Chandrasekhar (1931), Landau (1932) and the formalism developed by Tolman (1939), Oppenheimer, and Volkov (1939) predicted an upper limit on the NS mass in the range $M_{NS} = 0.7$–$3.4\ M_\odot$. The canonical mass, $M_{ch} \sim 1.4\ M_\odot$ is the critical mass above which the degenerate remnant core of a massive star or a WD loses gravitational stability and collapses into an NS. A more accurate parameterization of the Chandrasekhar mass limit (Chandrasekhar 1931) can be expressed as $M_{ch} = 5.83\ Y_e^2\ M_\odot$, where $Y_e = n_p/(n_p + n_n)$ is the electron fraction. Perfect neutron–proton equality ($n_p = n_n$) with $Y_e = 0.50$ gives the critical mass $M_{ch} \sim 1.47\ M_\odot$. However, the inclusion of more reasonable electron fractions $Y_e < 0.50$ gives smaller values for $M_{ch}$. The implication of the General Relativity (GR), correction of surface boundary pressure, and the pressure reduction due to non-ideal Coulomb interactions between $e^-$–$e^-$ repulsion, ion–ion repulsion, and $e^-$–ion attraction at high densities led to a decrease in the upper limit of $M_{ch}$.

Conversely, the electrons of the WD or the core of a massive star are not completely relativistic. This reduces the pressure leading to an increase in the amount of mass required to reach the gravitational potential to collapse the star. The entropy corrections and the effects of rotation also enhance the stability for additional mass. As a result, these corrections yield a higher upper limit for $M_{ch}$. The level of impact on the birth masses due to some of these competing effects is not well constrained as the details of the processes are not well understood. Thus, all of these facts suggest a wider mass range of $M_{ch} \sim 1.17$–$1.75\ M_\odot$. It seems that the measured masses refer to the effective gravitational masses, and not to the baryon mass ($M_{baryon}$) content. After applying the quadratic correction term $M_{baryon} - M_{grav} \approx 0.075\ M_{grav}^2$ (Timmes et al. 1996), one can obtain $M_{birth} \sim 1.08$–$1.57\ M_\odot$ as a viable range for gravitational mass of NSs, which is believed to encapsulate the range of NS birth masses. In addition, the amount of mass accreted ($M_{ac}$) onto an NS can be considered, which ranges $\Delta M_{ac} \approx 0.1-0.2$ $M_\odot$ (Kiziltan et al. 2013).

There is a large amount of literature that discusses the range of NS masses, and this problem still remains a subject of discussion. An earlier observation finds that NSs possibly predominantly range in $M_{NS} = 1.3–1.6$ $M_\odot$ (Finn 1994). A comprehensive observation of mass distribution of NS pulsars has been predicted with the narrow mass range of $\sim 1.38^{+0.10}_{-0.06}$ $M_\odot$ (Thorsett and Chakrabarty 1999). The maximum possible NS mass is mainly interesting due to the physical mechanism of its formation and gravitational stability, since the maximum NS mass exists near the limit of the small mass of stellar BHs (Fryer and Kalogera 2001; Rhoades and Ruffini 1974). The NS density gives a clear idea of the structure of matter that exists in a state of supernuclear density (Cook et al. 1994; Lattimer and Prakash 2004, 2007), which may also be responsible for the upper limit of NS mass. The observation of millisecond pulsars showed that several NSs consist of much higher mass than the canonical value of $M_{NS} = 1.4$ $M_\odot$ (Champion et al. 2008; Freire et al. 2008a,b; Ransom et al. 2005). The X-ray binaries in NSs also show systematic deviations from the canonical mass limit (Barziv et al.2001; Güver et al. 2010; Özel et al. 2009; Quaintrell et al. 2003; van Kerkwijk et al. 1995).

The EoSs for which it is assumed that the maximum NS mass $M_{NS-max} \leq 2.1$ $M_\odot$ (Gupta et al.2020; Landau 1932) are physically useless and do not fully agree well with recent observational results (Abbott et al. 2020a,c). The implied rigidity of EoSs largely excludes the presence of meson condensates and hyperons at supernuclear densities. Therefore, lower center densities, larger radii and thicker NS crusts are preferable for EoSs (Shapiro and Teukolsky 1983). The energy density–radius relation implied by Tolman (1939), combined with the causality limit, provides an analytical solution for an upper limit of the central density ($\rho_c$), $\rho_c M^2 = 1.53 \times 10^{16}$ $M^2_\odot$ g cm$^{-3}$. The lower limit of the maximum NS mass is $M_{NS} = 2.1$ $M_\odot$, the upper limit ($\rho_{max}$) of the NS central density is $\rho_{max} \leq 3.47 \times 10^{15}$ g cm$^{-3}$, which corresponds to $\approx 11\rho_s$ for the control saturation threshold $\sim 0.16$ $fm^{-3}$. Further, the exotic matter and Bose condensates significantly reduces the maximum mass of NSs. Consequently, the strict lower limit of the maximum NS mass $M_{NS-max} > 2.1$ $M_\odot$ excludes soft EoSs with extremely low density, which require the existence of exotic hadronic matter (Shapiro and Teukolsky 1983). The NSs with deconfined strange quark matter mostly have maximum predicted masses lower than $M_{NS-max} < 2.1$ $M_\odot$. Therefore, the EoSs with strange quark matter, which predict maximum masses less than $M_{NS-max} < 2.1$ $M_\odot$, can also be disregarded as useful configurations for NS matter.

Recently, using a gravitational wave detector, a much larger NS mass was observed in the NS-NS and NS-BH binary systems (Abbott et al. 2019, 2020a,b,c, 2021a). A number of other observations studies have predicted that the maximum mass of NS exists in the range $M_{NS-max} \approx 2.2–2.9$ $M_\odot$ (Bombaci 1996; Heiselberg and Pandharipande 2000; Kalogera and Baym 1996). Therefore, theoretical prediction using EoS (Shapiro and Teukolsky1983) to describe the highest NS mass in the previous paragraph is inconsistent with recent observational results. Furthermore, in GR, the maximum NS mass exists almost up to $M_{NS} \sim 3$ $M_\odot$. It is not clear whether the limit of GR is supported by rigid EoSs.

The maximum permissible NS mass is one of the most difficult questions in NS physics, and it is strongly related to the EoSs and the accepted theory of gravity. In other words, gravity and EoSs can together determine the maximum NS mass. The calculation of the exact value of the maximum NS mass strongly depends on the EoSs of neutron matter up to very high densities, $\rho_{high} \sim 5 \times 10^{15}$ g cm$^{-3}$. It is worth noting that the EoSs are obtained from exact numerical solutions of the quantum mechanical nuclear many-body problem using two- and three-particle potentials, which provide information in the region of the saturation density of nuclear matter, $\rho_{nucl} \sim 2.7 \times 10^{14}$ g cm$^{-3}$ (Wiringa et al. 1988). The EoSs can be considered up to a value of $\sim 2\rho_{nucl}$, and its expansion to $\sim 4\rho_{nucl}$ is possible. However, at much higher densities, the properties of matter remain uncertain, and the internal three-nucleon forces become significant given the hadronic properties, and ultimately the quark degrees of freedom become more relevant (Baym 1995). Increasing the matter density, the EoS approaches the sublight speed of sound, and at this moment the NS mass becomes equal to $M_{NS} \sim 3.2\ M_\odot$. For a compact star outside this mass limit, stronger short-range nuclear repulsive forces lead to an increase in the stiffness of the EoSs beyond the sublight speed limit. An upper limit on the mass of a compact star still exists in GR up to $M_{star} \approx 5.2\ M_\odot$, which takes into account spheres with a uniform density (Shapiro and Teukolsky 1983). The GR, along with the sublight speed limit, offers an upper limit for the maximum NS mass $M_{NS} \sim 3.2\ M_\odot$. The EoSs are influenced by taking into account nuclear processes, such as kaon and nucleon-nucleon scattering. These EoSs provide the mass range $M_{NS} = 1.5–2.2\ M_\odot$ as the lower limit of the maximum NS mass (Kalogera and Baym 1996; Thorsson et al. 1994). Although these lower bounds for the maximum NS mass are implied for many more realistic EoSs, it is still unclear if any of these values are realistic.

The first significant constraint for the EoS comes from the observation of two-solar-mass ($2\ M_\odot$) stars (Antoniadis et al. 2013). This means that the corresponding mass-radius curve must support sufficiently massive stars, $M_{max} > 2\ M_\odot$. This requires the stiff EoSs, which is combined with the fact that the high-density EoSs are quite soft (with $c_s^2 \lesssim 1/3$; $c_s$ is the speed of sound). A sharp cut-off of maximum NS mass in binary systems was studied using Gaussian mixture and Bayesian models (Alsing et al. 2018). In this model, it was shown that the maximum NS mass exists between $2.0\ M_\odot < M_{max} < 2.2\ M_\odot$ (68 % confidence) and $2.0\ M_\odot < M_{max} < 2.6\ M_\odot$ (90 % confidence). Cut-off proofs are robust to the choice of mass distribution model. The abrupt cut-off was interpreted as the maximum stable NS mass allowed by the EoSs of dense matter. The lower limit of the maximum speed of sound achievable inside the NS is $c_s^{max} > 0.63$ (99.8 % confidence), which excludes $c_s \lesssim c_s/\sqrt{3}$ at high significance.

On the basis of Einstein's theory of relativity, the principle of causality, and Le Chatelier's, it was established that the maximum mass of the equilibrium configuration of the NS cannot exceed $3.2\ M_\odot$ (Rhoades and Ruffini 1974). This limit is higher than the EoSs. The extremal principle, which follows from the theory of GR, was well applied when the EoSs of matter were unknown in a limited range of densities. The absolute maximum mass of NSs provides a decisive method for observing the difference between NSs and BHs. It was shown that the mass distribution of NSs goes beyond $2.5\ M_\odot$ taking into account flat and bimodal analysis. Indeed, the $M_{max}$ posterior is particularly sensitive to the inclusion of GW190814. For the flat model, the inferred maximum mass in the GW population of NSs shifts from $2.0^{+0.4}_{-0.3}\ M_\odot$ to $2.7^{+0.2}_{-0.2}\ M_\odot$. For the bimodal model, a comparable shift from $2.1^{+0.7}_{-0.3}$ $M_\odot$ to $2.7^{+0.7}_{-0.1}$ $M_\odot$ was observed. This difference could be indicative of a radio selection effect discussed in Galaudage et al. (2021).

Fattoyev et al. (2020) showed that in GW190814 by adjusting the existing energy density functional, it is possible to take into account the mass of NS 2.6 $M_\odot$, while satisfying the initial constraint on the tidal deformability of NS mass of 1.4 $M_\odot$, and reproducing the ground-state properties of finite nuclei. It was reported that the stiffening of the EoS required to support supermassive NS was inconsistent with either constraint obtained from energetic heavy-ion collisions or from the low deformability of medium-mass stars. Thus, it was suggested that the maximum NS mass cannot be significantly higher than the existing observational mass limit of 2.6 $M_\odot$. This indicate that such a compact object is likely to be the lightest BH ever discovered. However, , it was shown that the maximum NS mass is ~ 3 $M_\odot$ ($M_\odot \approx 2 \times 10^{33}$ g), and is limited only by causality and General Relativity (Nauenberg and Chapline 1973).

Theoretical and observational arguments suggest that stellar evolution may not produce BHs with masses less than ~5 $M_\odot$ (Bailyn et al. 1998; Farr et al.2011), and NSs are expected to have a maximum mass of ~3 $M_\odot$ (Kiziltan et al. 2013; Özel et al. 2012; Rhoades and Ruffini 1974). However, observation showed that the heaviest NS observed to date has a mass of $2.01^{+0.04}_{-0.04} M_\odot$ (Antoniadis et al. 2013). It has recently been stated that PSR J0740 ± 6620 may contain an NS of $2.14^{+0.10}_{-0.09}$ $M_\odot$, but systematic uncertainties in this measurement are still a matter of debate (Alsing et al.2018). The lack of observations of compact object in the 2-5 $M_\odot$ range to date hints at the existence of the so-called low-mass gap (Bailyn et al. 1998; Belczynski et al. 2012; Özel et al. 2010). From the analysis of the literature, it seems that the maximum mass of NSs observed by different groups are different. It can also be noticed that the observational results of the maximum mass of NSs does not fully agree with the mass determined from the EoSs and GR.

Let us discuss the constraint of the maximum mass of NSs determined from the EoSs and GR. The properties of NSs are determined by the nature of the matter that they contain. These properties can be constrained by measurements of the star size and mass. The stringent constraints on NS radii by combining multimessenger observations of the binary NS merger GW170817 with nuclear theory that best accounts for density-dependent uncertainties in the EoSs. The EoSs are limited by chiral effective field theory and are marginalized from them by observations of gravitational waves. Combining this with the electromagnetic observations of the merger remnant, which suggest the presence of a short-lived hypermassive NS, it was found that the radius of the NS is $11.0^{+0.9}_{-0.6}$ and the mass $M_{\rm NS}$ = 1.4 $M_\odot$ (Capano et al. 2020). The EoSs extended to higher densities do not do a low-energy extension of the effective field theory in general, while ensuring that the speed of sound is less than the speed of light and that the EoSs support an NS with a mass of 2 $M_\odot$. The multimessenger observations of GW170817 were used to constrain these EoSs to ensure that they are consistent with: (1) the detected gravitational waves during the inspiral; (2) the production of a post-merger remnant that does not immediately collapse to a BH; and (3) the constraints that the energetics of the GRB and kilonova impose on the maximum NS mass. The higher the mass of the heaviest observable NS, the more restricted the EoSs. In addition, the maximum densities in NSs are limited: stiff EoSs with large polytropic exponents have lower maximal densities, which are strongly constrained by causality. Softer EoSs tend to have larger central densities. For $M_{NS} = 1.97\ M_\odot$, the central densities are as high as $\approx 8.3\ \rho_0$, and for $M_{NS} = 2.4\ M_\odot$ the densities reach to $\approx 5.8\ \rho_0$.

The millisecond variability in the X-ray flux of several low-mass X-ray binaries (LMXBs) NSs was interpreted within a general relativistic framework of accretion, in which a cutoff frequency was expected. This testifies the severe constraints on the EoSs of matter at supranuclear densities. The reported maximum frequency (1.15±0.02 kHz) of quasiperiodic oscillations observed in sources as diverse as Sco X-1 and 4U 1728-34 (Kluźniak 1998) would imply that the neutron star masses in these LMXBs are $M_{NS} >1.9\ M_\odot$, and hence several EoSs would be excluded. Consequently, the above results showed that the NSs can exist in the range of maximum mass $M_{max}$= 1.5–3.2 $M_\odot$ (Kiziltan et al. 2013). The existence of the highest NS mass and lowest BH mass becomes critical for considering the "mass gap" (Abbott et al. 2020a,c; Belczynski et al. 2012; Kreidberg et al. 2012; Littenberg et al. 2015; Mandel et al. 2015; Ozel et al. 2010, 2012 Wyrzykowski and Mandel 2020; Yang et al. 2020; Zevin et al. 2020).

## 3. MASS DISTRIBUTION OF BINARY COLLAPSARS AND "MASS GAP"

### 3.1. OBSERVATIONAL LIMIT FOR "MASS GAP" AND COMPACT OBJECTS FILLING THE "MASS GAP" REGION

Note that the "m-gap" region is considered between the confirmed observations of the minimum mass limit of astrophysical BHs and the maximum NSs masses. In the mass distribution, this region exists at $M_{\text{m-gap}} \sim 2-5\ M_\odot$ (Kreidberg et al. 2012; Özel et al. 2010, 2012). However, Thompson et al. (2019) reported a giant star J05215658 of a mass $M_{star} \sim 3.2^{+1.0}_{-1.0}\ M_\odot$, and a non-interacting low-mass BH companion with a mass of $\sim 3.3^{+2.8}_{-0.7}\ M_\odot$. It was claimed that the compact object with such a mass is located in the "m-gap" between NSs and BHs. One of the highest NS mass was observed by Antoniadis et al. (2013) and Cromartie et al. (2020) with the masses $M_{NS} \sim 2.01^{+0.04}_{-0.04}\ M_\odot$ and $\sim 2.14^{+0.01}_{-0.09}\ M_\odot$, respectively. The uncertainty range nearly spans from the predicted theoretical maximum NS mass $M_{NS} \simeq 2.5\ M_\odot$ (Lattimer2012) to the lowest well-measured BH masses $M_{BH}$ = 5–6 $M_\odot$ (Farr et al. 2011; Özel et al. 2010). A model proposed by Kochanek (2014) showed a lower mass limit for BH formation with the gravitational mass $M_{BH} \sim 4\ M_\odot$, and such a mass limit of the BH was observationally detected in XTE J1650–500 (Orosz et al. 2004). Heida et al. (2017) have claimed that GX 339–4 with 2.3 $M_\odot < M_{BH} < 9.5\ M_\odot$ is possibly the first BH to fall in the "m-gap" $M_{\text{m-gap}}$ = 2–6 $M_\odot$. Moreover, it was reported by Orosz et al. (1998) that the mass of the CBS in 4U 1543–47 is likely to be $\sim 3\ M_\odot$ which critically exceed the mass of NS, and therefore, this compact object was claimed to be a BH. In recent observation, a BH candidate of the mass $M_{BH} \sim 3\ M_\odot$ as a binary companion to V723 Mon was discovered (Jayasinghe et al. 2021).

The results showed firm evidence that the target star in a binary system with a non-luminous object has the minimum mass $M_{min} = 4.36^{+0.41}_{-0.41}\ M_\odot$ (Giesers et al. 2018). This object supposed to be degenerate because it is invisible, and the minimum mass is significantly higher than the Chandrasekhar limit ($\sim$1.4 $M_\odot$) (Chandrasekhar 1931). The masses of several NSs binary pulsar was measured and assumed that the NS masses are uniformly distributed between lower ($M_{NSl}$) and upper mass ($M_{NSu}$) bounds (Finn 1994):

$$1.01 < M_{\text{NS-}l}/M_\odot < 1.34, \quad 1.43 < M_{\text{NS-}u}/M_\odot < 1.64.$$

These limits provide observational support to NS formation mechanism, which suggests the mass range of $1.3 < M_{\text{NS}}/M_\odot < 1.6$. Most likely, the degenerate object exceeded the Tolman–Oppenheimer–Volkoff limit, according to which all objects collapse in BH above $M \sim 3\ M_\odot$ (Bombaci1996). It is noted that the estimate of the mass of the dark companion weakly depends on the mass of the target star, within a reasonable error, for example, for the unrealistic case of a target star with $0.2\ M_\odot$, the minimum companion should be above $3\ M_\odot$.

The results of studies by Giesers et al. (2018); Jayasinghe et al. (2021); Orosz et al. (1998) showed the presence of few low-mass BHs ($M_{\text{BH}} < 5\ M_\odot$) and a most prominent likelihood distribution around $M_{\text{BH}} = 7$–$8\ M_\odot$ with a sharp decline beyond $10\ M_\odot$ (Özel et al. 2012). This distribution is quite different from the "high-mass" BH binary systems, and provides interesting constraints on the supernovae and binary evolution processes that create BH soft X-ray transients. However, it is assumed that the gap of BHs with masses less than $M_{\text{BH}} < 5\ M_\odot$ is expected to be associated with supernova explosions and lead to the formation of BHs that fill the "m-gap" (Fryer et al. 2002; Fryer and Kalogera 2001). A calculation showed that supernova explosions typically generate a continuous distribution of BH masses, and mass decays exponentially to the higher mass end in the distribution. The continuity of masses is primarily a consequence of the relatively gradual dependence of explosion energies on the masses of the progenitors. It is assumed that the energies of the explosion can be comparable and/or less than the binding energy of stellar envelopes (Fryer et al. 2002; Fryer and Kalogera 2001). Although the creation of an "m-gap" in the mass distribution is impossible with the current understanding of the energy of a supernova. It has been suggested that this can be achieved under the special assumption that the explosion energy has a stepwise dependence on the progenitor mass and that it drops to zero for stars more massive than $\sim 25\ M_\odot$ (Fryer and Kalogera 2001). This bimodality in the energies of explosions that form NSs compared to BHs should become apparent in the large sample of supernovae expected in current research that are sensitive to faint CCSNs (Burrows and Vartanyan 2021).

The heaviest NSs and lightest BHs are expected to be formed as a result of stellar evolution with a mass range of $2.2\ M_\odot < M < 5\ M_\odot$, which is largely unpopulated. The compact objects found in the "m-gap" are probably associated with the CCSN explosion, which is well described in the review article by Burrows & Vartanyan (2021). In CCSNs, it was suggested that if stellar-mass BHs are born as a result of collapse and NSs have a maximum mass just above about $2.0\ M_\odot$, then there may be an "m-gap" between them. The object with a mass of $M = 2.6\ M_\odot$ was recently discovered in the binary merger of GW190814 through gravitational waves by LIGO/Virgo (Abbott et al. 2020c). In this observation, a compact binary merger involving a BH with the mass $M_{\text{BH}} = 22.2$–$24.3\ M_\odot$ and a compact object with a mass of $2.50$–$2.67\ M_\odot$ was reported. In this system, the source has the most unequal mass ratio measured so far with gravitational waves, and the secondary component with a mass of $2.50$–$2.67\ M_\odot$ is either the lightest BH or the heaviest NS ever discovered in CBSs. The astrophysical models predict that CBSs with mass ratios similar to this event can be formed through several mechanisms, but are unlikely to be formed in globular clusters (Abbott et al. 2020c). However, the combination of mass ratio, component masses, and the incidental merger rate for GW190814 detected by gravitational wave describes the mass distribution of compact binaries. Abbott et al. (2020c) reported that comparison of the secondary mass and several current estimates of the maximum NS mass suggests that GW190814 is unlikely to be formed from NS-BH coalescence. Nevertheless, this event sheds light on a new concept on the mass distribution of compact objects at the interface between known NS and BH masses. It was explained that the BHs in the "m-gap" naturally gather through mergers and accretion in the active galactic nucleus (AGN), and subsequently can participate in additional mergers (Yang et al. 2020). The lighter object of GW190814, with a mass of 2.6 $M_\odot$, could have grown in an AGN disk through accretion. However, surprisingly, the high total mass of 3.4 $M_\odot$ observed in the NS merger of GW190425 (Abbott et al. 2020a) may also be formed due to accretion in an AGN (Yang et al. 2020).

In the subsequent investigation, a component mass range from 1.12 to 2.52 $M_\odot$ was observed in the GW190425 event (Abbott et al. 2020a). This value becomes reasonable in the range 1.46–1.87 $M_\odot$ if the dimensionless spin values of the components can be limited to be less than 0.05. These masses are consistent with the individual binary components of NSs. However, both the source-frame chirp mass $1.44^{+0.02}_{-0.02}$ $M_\odot$ and the total mass $3.4^{+0.1}_{-0.3}$ $M_\odot$ of this binary are significantly larger than those of any other known binary NSs. On the basis of gravitational wave analysis, it cannot be rejected that both binary components of the system are BHs. Subsequently, it was found that the gravitational waves of the binary system GW170817 arise as a result of the merger of two compact objects in the NS mass range (Abbott et al. 2020b). The gravitational wave results indicate the possibility that one or both objects are low-mass BHs. If the coalescing BH after the merger rotates slowly, then the maximum baryon mass of nonrotating NS contains the mass $M_{NS} = 3.05$ $M_\odot$. In this case, the existence of the EoSs was supposed to be excluded. A tighter NS mass limit was proposed to be $M_{NS} = 2.67$ $M_\odot$, if the merging leads to a hypermassive NS (Abbott et al. 2020b). Using various statistical methods, the mass of the PSR J0740+6020 system, with the pulsar mass $2.08^{+0.07}_{-0.07}$ $M_\odot$ was determined by the relativistic Shapiro time delay with 68.3% credibility (Fonseca et al. 2021). This mass also falls in the "m-gap" region.

It was reported that the original masses of the components of GW200105 and GW200115 arose from the merging of NS and BH (Abbott et al.2021a). In this work, their primary masses were reported to be $M_1 = 8.9^{+1.2}_{-1.5}$ $M_\odot$ and $5.7^{+1.8}_{-2.1}$ $M_\odot$, respectively, which is consistent with the predictions of BH masses in population synthesis models for NSs and BHs. The secondary masses involved in the GW200105 and GW200115 events were detected to be $M_2 = 1.9^{+0.3}_{-0.2}$ $M_\odot$ and $1.5^{+0.7}_{-0.3}$ $M_\odot$, respectively. The results are consistent with the observed NS mass distribution in the Milky Way, as well as predictions of population synthesis for secondary masses in NS and BH merging. In the work (Abbott et al. 2021a), no firm evidence was reported describing that the mass of secondary compact objects belongs to NSs. In the observation, it was pointed out that the masses are consistent with either a binary BH merger or the merger of NS and BH. Moreover, the comparisons of the secondary masses to the maximum allowed NS mass yield a probability $p(M_2 \leq M_{max})$ of 89%–96% and 87%–98% for the secondaries in GW200105 and GW200115, respectively, which is compatible with the mass of NSs.

The upper bound of the NS mass $M_{NS} = 2.9$ $M_\odot$ was established regarding the EoS, valid up to $2\rho_{nm}$ of the saturation density (Kalogera and Baym 1996). Reliable evidence which confirm that a compact object is a BH can be obtained if the mass limit exceeds the value $M_{\text{NS-max}} = 2.9\ M_\odot$. According to the EoS, the lower BH mass limits are in the range $M_{\text{BH}} = 3.1$–$6\ M_\odot$. It is quite probable that more compact objects can be identified as BHs since the value of $M_{\text{max}} = 2.9\ M_\odot$ is substantially lower than $3.1\ M_\odot$. Gelino and Harrison (2003) have reported the smallest stellar BH (J0422+32) with the mass $M_{\text{BH}} = 3.97 \pm 0.95\ M_\odot$. Thus, it was claimed that J0422+32 contains a stellar BH with the lowest mass, and it falls in the "m-gap" range $M_{\text{m-gap}} = 3$–$5\ M_\odot$.

However, it is argued that the formation of GW190814 at any measurable rate requires a supernova engine model that operates on longer time scales, so that a proto-compact object can undergo significant accretion just before the explosion (Zevin et al. 2020). This is a hint that if GW190814 (Abbott et al. 2020c) is the result of the evolution of a massive binary star, then the "m-gap" between NSs and BHs may be narrower or absent at all.

Wyrzykowski and Mandel (2020) reported the possible existence of an "m-gap" between the highest NS mass and lowest BH mass. It was argued that the heaviest NSs should have the mass $M_{\text{NS}} = 2\ M_\odot$, while mass of the lightest BHs should be about $M_{\text{BH}} = 5\ M_\odot$. The "m-gap" between the NS and BH masses in the range $M_{\text{m-gap}} = 2$–$5\ M_\odot$ was described to be inconsistent with observational data, unless BH receives natal kicks above $> 20-80\ \text{km s}^{-1}$ (Wyrzykowski and Mandel 2020). In this work, eight candidates for objects with masses within the assumed mass gap were identified, including a spectacular multi-peak parallax event with a mass of $2.4^{+1.9}_{-1.3}\ M_\odot$. In these circumstances, the "gap" in the mass distribution has become a serious problem for understanding the formation of CBSs. The mass distribution of NSs and BHs with the "m-gap" between these compact objects was proposed to predict by combining the results of stellar modeling with hydrodynamic modeling of supernovae (Belczynski et al. 2012). However, determining the highest mass of NSs and the lowest mass of BHs is a challenging task due to difficulty in obtaining the precise gravitational potential and deterministic value of EoSs.

## 3.2. STATISTICAL DISTRIBUTION OF MASSIVE NEUTRON STARS AND LOW MASS BLACK HOLES

Littenberg et al. (2015) demonstrated the quantitative capabilities of LIGO/Virgo to distinguish between the mass of NSs and BHs from a large number of sources. The simulated population of compact objects showed the smaller component with the mass equivalent to NSs $M_{\text{NS}} \lesssim 1.5\ M_\odot$, while the population of compact mass of binary BHs $M_{\text{BH}} \gtrsim 6 M_\odot$ was also confirmed. High-mass NSs ($2 < M_{\text{Ns}} < 3\ M_\odot$) are often consistent with low-mass BHs ($M_{\text{NS}} < 5\ M_\odot$), and it is posing a challenge for determining the maximum NS mass and solving the "m-gap" problem (Littenberg et al.2015). Özel et al. (2012) found that the distribution of NS masses in non-recycled eclipsing high-mass binaries as well as of slow pulsars consist of birth masses with a mean value of $1.28\ M_\odot$, and a dispersion of $0.24\ M_\odot$. These values are considered to correspond to the formation of NSs in the CCSN explosions. Binary NSs have been reported to have a very narrow mass distribution, peaking the value at $1.33\ M_\odot$ with a dispersion of $0.05\ M_\odot$ (Fig. 1a). Such a narrow distribution is difficult to explain within the framework of the current understanding of the mechanisms of NS formation. One of the possible ways to obtain a narrow distribution is via electron capture supernovae in ONeMg white dwarfs (Podsiadlowski et al. 2005). The emergence of such a supernova occurs at a certain density threshold, which corresponds to a pre-collapse mass of the white dwarf in the narrow range 1.36–1.38 $M_\odot$ for different temperatures and compositions. The mean mass distribution of recycled NSs is 1.48 $M_\odot$ with a dispersion of 0.2 $M_\odot$. A small fraction of recycled NSs in the incidental distribution has masses that exceed $M_{NS} \sim 2\ M_\odot$, and it was suggested that only some of these NSs overcome the mass threshold to form low-mass BHs.

In a subsequent work, Özel et al. (2010) have performed the measurements of the dynamic mass of 16 BHs in transient low-mass X-ray binaries in order to infer the mass distribution of stellar BHs in the parent population. The observational results were best described by a narrow mass distribution at $7.8^{+1.2}_{-1.2}\ M_\odot$. A selection effect related to the choice of targets for optical follow-ups was identified, which led to a flux-limited sample. However, it is assumed that the selection effect does not introduce a bias for the observed distribution (Özel et al. 2010), and it cannot explain the absence of BHs in the "m-gap" region $M_{m\text{-}gap} = 2\text{–}5\ M_\odot$. It was argued in Özel et al. (2010) that the rapid decline at the high-mass end in the estimated distribution may be associated to a specific evolutionary mechanism followed by low-mass X-ray binaries (Fig. 1b). Mass transfer in the low-mass X-ray binary phase requires small orbital distance between the BH and the low-mass companion. The common envelope phase, which is invoked to reduce sufficiently the orbital separation of the binary prior to the supernova, is thought to lead to the expulsion of the hydrogen envelope of the pre-supernova star, leaving behind a bare helium core. Furthermore, winds from the resulting helium cores in the Wolf–Rayet phase are expected to lead to further mass loss, albeit at rates that are highly uncertain. As a result, it is possible that the BH masses in contact binaries are mostly persevered near the mass of 10 $M_\odot$. This evolutionary path could provide a natural explanation for the rapid decline of the inferred mass function at the high-mass end.

It is predicted that the observed mass distribution can differ from the distribution of birth masses (Fragos and McClintock 2015). This is due to mass transfer in the X-ray binary phase, which could lead to the fact that BHs evolved from the masses they had at the time of birth. The least massive BHs are predominantly born with low-mass companions ($M_c \lesssim 2\ M_\odot$) and, therefore, cannot accrete a significant amount of matter during the lifetime of the binary system (Fragos and McClintock 2015).

If the companions are more massive, $M_c > 5\ M_\odot$, at the beginning of the mass transfer phase, the BHs are born with the masses $M_{BH} \gtrsim 5\ M_\odot$. As a result, the accretion in the binary phase is not likely to bridge the gap of mass distribution between $M_{m\text{-}gap} = 2\text{–}5\ M_\odot$.

Moreover, it was suggested that BH evaporation in braneworld gravity models can lead to a deficiency of low-mass BHs in the steady-state population. The "m-gap" in the distribution could be created in this context because the rate of evaporation in braneworld gravity increases rapidly with decreasing BH mass (Postnov and Cherepashchuk2003). However, recent constraints on the rate of evaporation obtained using the current population of BHs exclude this possibility (Rhoades and Ruffini 1974).

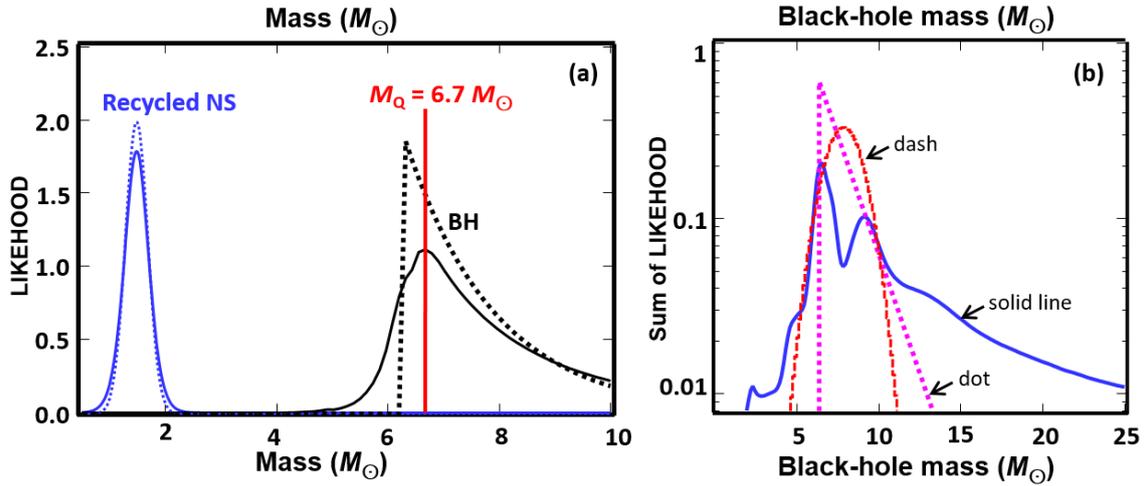

**Fig. 1.** *Left* (a) – The incidental mass distributions of recycled population of massive NSs and low-mass black holes; the dotted lines correspond to maximum likelihood values (Özel et al. 2012). The mass of NSs in the mass distribution is populated differently: $M = 1.33\ M_\odot$, $\sigma = 0.05\ M_\odot$ – double NSs, $M = 1.28\ M_\odot$, $\sigma = 0.24\ M_\odot$ – NSs near their birth masses, and $M = 1.48\ M_\odot$, $\sigma = 0.20\ M_\odot$ – recycled NSs. An exponential distribution was used for BHs with a low mass cut-off at $M_c = 6.32\ M_\odot$ and a scale of $M_{scale} = 1.61\ M_\odot$; the solid lines represent the weighted mass distributions for each population, and the dotted line represents the fitting slope. The red line in the mass distribution at $M_Q = 6.7\ M_\odot$ was calculated using the concept of gravidynamics (Sokolov and Zharykov 1993). *Right* (b) – Solid line represents the sum of likelihood masses of 16 BHs in low-mass X-ray binaries (Özel et al. 2010). The shape of their sum is artificial at the high mass end, which is caused by high-mass wings of the individual likelihoods, the dashed and dotted lines show the exponential and Gaussian distributions, respectively, with best fitting parameters.

Sokolov and Zharykov (1993) suggested that in gravidynamics and within the framework of GR, one can consider a spherically-symmetric configuration with the limited EoS $P_Q = 1/3(\varepsilon - 4B)$, where density increases towards the center of compact mass. This model assumed the existence of stable configuration of quark star with quark-gluon plasma that includes all possible quark flavors. The total mass of such a compact object with a radius of ~ 10 km consists of a peak mass of about 6.7 $M_\odot$. This peak, including the mass distribution of NSs and BHs, is highlighted in Fig. 1a.

A smaller value of BH mass ($M_{BH} < 4$–$5\ M_\odot$) was predicted for the compact objects of GRO J0422+32 and 4U1543−47 (Fig. 2a), which eliminates the "m-gap" identified by Ozel et al. (2010). However, precise measurements are suggested to confirm this claim. The results of "m-gap" is considered correctly and consistent only if the GRO J0422+32 system is excluded from the analysis. However, if GRO J0422+32, and perhaps 4U1543−47 does fall into the supposed "m-gap", the basic features of the mass distribution of BH soft X-ray transients remain.

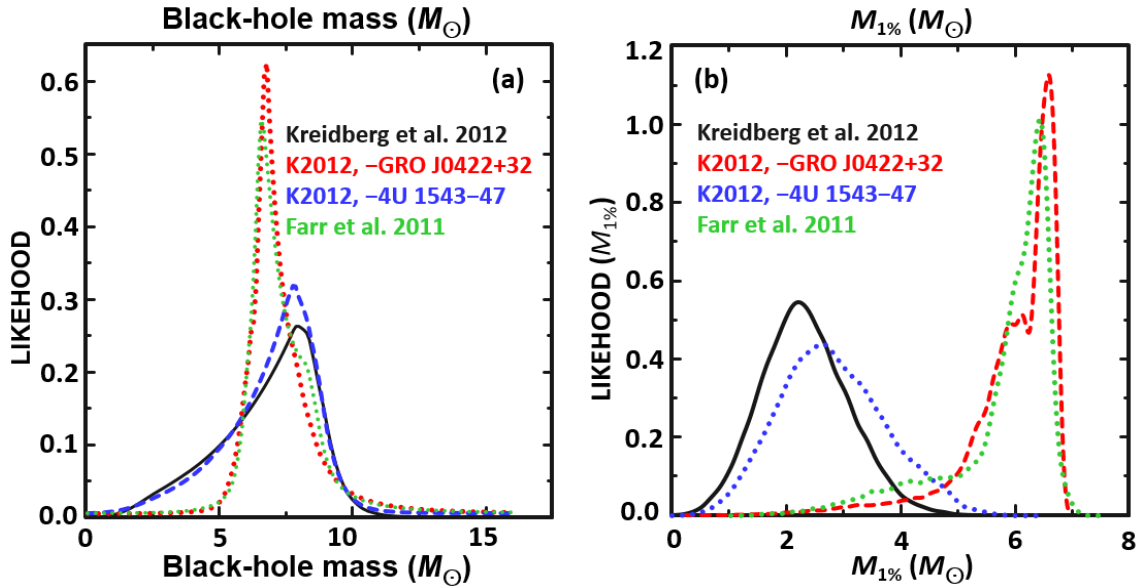

**Fig. 2.** *Left* (a) – Probabilistic likelihood of mass in different versions of BH distributions calculated by the power-law model (Kreidberg et al. 2012); black curve – using parameters of 16 BH binary systems; red curve – parameters of 16 binary BH systems of (Kreidberg et al. 2012), excluding GRO J0422+32; blue curve – system parameters and exclusion of 4U 1543−47; green curve – power-law analysis with the using of parameters of 16 binary BH systems of Farr et al. (2011); *Right* (b) – Probability distribution of the 1% mass quantile, $M_{1\%}$, in the power-law model implied by the analysis with system parameters (black curve); red curve – parameters with the exception of GRO J0422+32; blue curve – parameters with the exception of 4U 1543−47; and green curve − parameters from Farr et al. (2011); minimum BH masses extend through the gap ($M_{1\%} \lesssim 4\ M_\odot$); minimum BH mass is equivalent to the analysis from Farr et al. (2011) with the elimination of GRO J0422+32.

Farr et al. (2011) described the mass distribution of stellar-mass BHs in X-ray binary systems using Bayesian analysis. In this analysis, a sample of 15 low-mass, Roche lobe filling systems and a sample of 20 systems containing the 15 low-mass systems and 5 high-mass, wind-fed X-ray binaries was considered. The Markov Chain Monte Carlo calculations methods was used to sample the posterior distributions of the parameters derived from the data for five parametric and five non-parametric models for the mass distribution. This study provided evidence for an "m-gap" between the most massive NSs and the least massive BHs. The most favorable power-law model for the low-mass systems, gives a BH mass distribution whose 1% quantile lies above 4.3 $M_\odot$ with 90% confidence. However, for the combined sample of systems, the most favorable, exponential model gives a BH mass distribution whose 1% quantile lies above 4.5 $M_\odot$ with 90% confidence. It was concluded from the results of Farr et al. (2011) that BH masses provide evidence for an "m-gap" between the maximum NS mass and the lower bound of BH masses. It was established that both the low-mass and combined population require the presence of a gap between 3 and 4–4.5 $M_\odot$ (Bailyn et al.1998; Farr et al. 2011; Özel et al. 2010).

Two possible reasons for the existence of such a gap were proposed by Fryer (2002), who indicated a step-like dependence of supernova energy on progenitor mass or selection biases. The core collapse in massive stars may be associated with the dependence of supernova energy on progenitor mass. Selection biases can occur because the X-ray binaries with low mass BHs are more likely to be persistently Roche lobe overflowing, which prevent dynamical mass measurements. It was concluded that the existence of bias is not associated with hard evidence for considering the "m-gap" (Özel et al. 2010). However, it was argued that the observed stable X-ray sources was not belong to NSs, and insufficient to populate the $M_{\text{m-gap}} = 2–5\ M_\odot$ region of any BH mass distribution, which increases towards low masses in the distribution. It has been suggested that models of population synthesis involving binary evolution and transitional behavior (Fragos et al. 2008, 2009) may help to define the problem of the "m-gap". The shapes of the mass distribution of NSs and BHs are distinctly different in the distribution function (Fig. 1a). This is similarly true for the mass distribution involving the NS–NS and NS–WD binaries (Kiziltan et al. 2013). The analysis proposed by Kiziltan et al. (2013) showed consistent corresponding peaks at $M_{\text{NS}} = 1.33\ M_\odot$ and $1.55\ M_\odot$ of NSs in the NS–NS and NS–WD binaries, suggesting significant mass accretion $\Delta M_{\text{acc}} \approx 0.22\ M_\odot$ during the spin-up phase. The width of the mass distribution implied by binary NS systems indicates a narrow initial mass function, while the assumed mass range is much wider for NSs that have experienced recycling. The masses of NSs in NS–WD binaries showed evidence of skewness with a heavy tail on the high-mass end (Fig. 3a) and exponential tendency for lower mass distribution. However, the NS–NS binaries imply a narrow almost Gussian symmetric mass distribution (Fig. 3b).

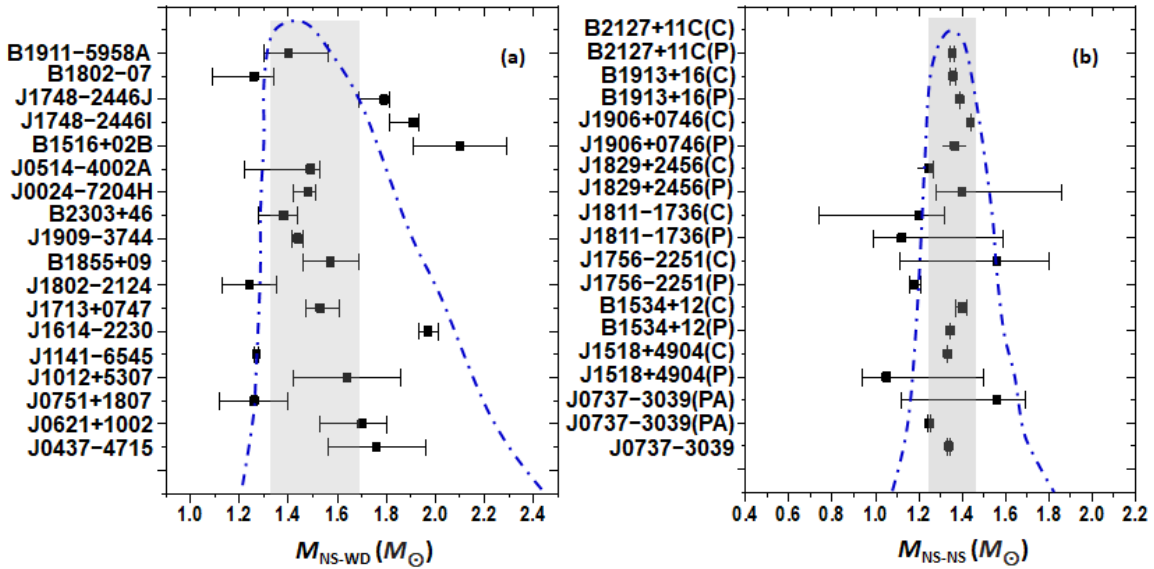

**Fig. 3.** *Left* (a) – Measured values of mass of radio pulsars of NS–WD, and *Right* (b) – NS–NS binaries; vertical band of regions are the peak values of the underlying mass distribution for NS-WD ($M_{\text{NS-WD}} = 1.55\ M_\odot$), and double NSs ($M_{\text{NS}} = 1.33\ M_\odot$) binaries systems; mass distribution for NS-WD is wider and NS-NS binaries is narrower (Kiziltan et al.2013); the light-dark band in these figures represents 80-90 % of mass distribution.

The populations of the binaries objects show no evidence of a strong truncation limit on either end. A mass cutoff at ~2.1 $M_\odot$ for NSs with WD companions established a lower bound for the maximum NS mass. This rules out the majority of strange quark and soft EoS models as viable configurations for NS matter. The lack of truncation close to the maximum mass cutoff along with the skewed nature of the inferred mass distribution both enforce the suggestion that the 2.1 $M_\odot$ limit is set by evolutionary constraints rather than nuclear physics or GR, and the existence of rare supermassive NSs is possible that can occupy the "m-gap". This has important consequences that are associated with the stochastic nature of evolutionary processes. A long-term gravitationally stable accretion naturally produces a wider distribution for NSs in NS–WD binaries (Fig. 3a). This, along with the lack of a strong truncation, indicates that, in particular, the high-mass end of the NS mass distribution is driven by evolutionary constraints. As a result, this rules out the possibility that an upper mass limit is set by the EoSs or GR for NSs with masses $M_{NS} \lesssim 2.1 \, M_\odot$. Therefore, the $M_{NS} = 2.1 \, M_\odot$ upper mass limit implied by NSs in NS–WD binary should be considered a minimum safe limit to the maximum NS mass rather than an absolute upper limit to NS masses. The heavy tail of the mass distribution of NSs in NS–WD binary (Fig. 3a) favors the possibility that at least some of these pulsars are born as massive NSs. The implications of these results suggest a prediction of the upper limit of the mass of the NSs, which may help to determine the "m-gap" in the lower part of the distribution function, as discussed in (Bailyn et al. 1998; Farr et al. 2011; Özel et al. 2010). The mass distribution in the NS–NS and NS–WD binaries may also hint to the correct shape of the mass distribution proposed by Özel et al. (2012) (Fig. 1) and the problem of "m-gap" in the distribution, which can be discussed in the context of binary NS merger.

It was observed that the binary NS mergers predominantly produce BH remnants of mass $M_{BH} \sim$ 3–4 $M_\odot$, thus populating the low "m-gap" between NSs and stellar-mass BHs (Gupta et al. 2020). If the low-mass BHs are in dense astrophysical environments, mass segregation could lead to "second-generation" compact binaries merging within the Hubble time. The astrophysical considerations suggest the possible existence of an "m-gap" between the heaviest NSs and the lightest stellar-origin BHs (Bailyn et al. 1998; Özel et al. 2010). The gap can be related to the selection effect (Farr et al. 2011), so it is important to verify whether BHs populating the "m-gap" exist in nature. In this aspect, the GW observations present the orthogonal selection effects compared to electromagnetic probes, thus offering a promising opportunity to settle this issue. Using the increasing number of GW probe, it becomes easier to determine whether the "m-gap" is populated, and consequently to set constraints on the astrophysical mechanisms by which the "m-gap" is populated (Littenberg et al.2015; Mandel et al.2015).

Understanding the existence of compact objects in the "m-gap" has important astrophysical implications. It was reported that the stellar collapse can only produce BHs with masses $M_{BH} \lesssim 5 \, M_\odot$ if the explosions are driven by instabilities that develop over timescales $\gtrsim$ 200 ms (Belczynski et al. 2012). In the mass distribution one can find a predicted gap if these instabilities develop on shorter timescales. There are several arguments indicating that the first binary NS merger GW170817 must have produced a hypermassive NS (Abbott et al. 2020b; Margalit and Metzger2017), which eventually collapsed to a BH in the low "m-gap". The total mass of the GW190425 binary is significantly larger than the mass of galactic double NS binaries. On this basis, it cannot be ruled out that the possibility of one or both binary components exist in BHs (Abbott et al.2020b; Margalit and Metzger 2017). These two observed events suggest that the ratio of NS–NS merger remnants to NSs in a Milky Way equivalent galaxy should be κ ∼ 0.01. This implies the existence of a population of low-mass BHs in merging compact binaries, which can be probed with the third-generation GW detectors. The inverse problem is also intriguing, in which measuring the relative abundance of NS mergers and low "m-gap" BH mergers allow to infer the typical number of NS mergers occurring in a galaxy during its cosmic lifetime. In the model, it is assumed that all merger remnants are retained inside the cluster and remain available to form 2g objects. Both natal and merger kicks might decrease the available number of low-mass BHs in clusters. The events of GW151226 and GW170608 (Abbott et al. 2019) hint at the existence of some BHs in the low "m-gap" $M_{\text{m-gap}}$ = 2–5 $M_\odot$.

BHs also can have a population of selected mass value, and peak in the mass distribution of these relativistic objects is close to $M_{\text{BH}} \sim 6.7$ $M_\odot$ for low-mass BHs (Özel et al. 2010). This is considered the smallest gravitational mass for the formation of a BH candidate. Different BH masses and peak in the distribution more probably reflect selection effects related with evolution of CBSs, in which such objects are observed. This suggests the possibility of mass distribution of NSs and BHs in compact binaries (Fig. 1a). The mass of NSs and BHs with "m-gap" can be explored in the context of relation between supernovae and GRBs. The solution of this problem can also be approached by the CCSN explosion mechanism. In particular, polarized radiation of long GRBs, black body spectrum, and other observational properties can be explained by collapsar properties (Sokolov 2015). The mass distribution of compact remnants and "m-gap" in the distribution of binary systems have been extensively studied, and this is highlighted above. However, it was found that there are lots of distinct and contradictory observational results and models, and therefore the problem is still open for investigation.

In the GR solution, the mass distribution of compact objects, such as NSs and BHs in CBSs, is considered as a continuum. However, a significant gap in the mass distribution between BHs and NSs was observed by Ozel et al. (2012) (Fig. 1a). The early evidence of the significant "m-gap" of ~ 3 $M_\odot$ between NSs and BHs was observed by Bailyn et al. (1998), Kalogera and Baym (1996). The masses of compact objects in X-ray binary systems with a low-mass optical companion star was listed by Ozel et al. (2010) and Kreidberg et al. (2012). The parameters of these binary systems have smallest measurement uncertainties, and the separation of mass distribution between NSs and BHs is noticeable. In these binary compact objects, the BH candidate mass is centered at near the maximum height of the peak located at $M_c$ = 6.32 $M_\odot$ (Fig. 1a), proposed by Ozel et al. (2010). This cut-off mass in the exponential distribution is well above the theoretical value, indicating a significant "m-gap" between NSs and BHs. The exponent mass scale $M_{\text{scale}}$ = 1.61 $M_\odot$ in mass distribution is significantly smaller than the expected theoretical value, which indicate moderately narrow mass distribution of BHs. The shape of mass distribution of BHs in the high mass end is artificial because it is related to the high mass wings of individual BHs.

The probability of BH mass distribution is calculated by using the power law model proposed by Kreidberg et al. (2012) and the results are shown in Fig. 2. However, our results are mainly consistent with those consequences, which were obtained by Özel et al. (2010), showing the peak of mass at $M$ = 8 ± 1.0 $M_\odot$ in the distribution function. Özel et al. (2010) studied 16 low mass X-ray binary objects containing BHs. The parametrized probability density distribution for 16 BHs in low massive X-ray binaries was obtained by considering the exponential function. The results showed a sharp break in the low mass end and sharp decline towards large masses of compact objects. The sharp decline of BHs mass does not agree with the theoretical predictions and the sizable gap between NSs and BHs is also not understood (Özel et al. 2010). The comprehensive highlights of the "m-gap" and distribution of the masses of NSs and BHs are thoroughly presented in the above sections. The properties of a compact object with the peak gravitation mass at $M_{\text{peak}}$ = 6.7 $M_\odot$ shown in Fig. 1a are a physically unresolved issue. However, with the consideration of GR, such a compact object is well described by the BH candidate (Yagi and Stein 2016). The calculation of probability distributions of compact gravitational objects was tested by four different groups (Farr et al. 2011; Kreidberg et al. 2012; Özel et al. 2010, 2012; Petrov et al. 2014), and it was however confirmed that the peak in mass distributions for BH candidate does not exist.

Kreidberg et al. (2012) reported that the existence of BHs in the "m-gap" is not possible. In the mass distribution the authors attempted to shift all 16 low massive X-ray binaries by only one 4U 1543-47 system in the region of NSs with the mass not more than $M_{NS} \approx 2.1\ M_\odot$ (Özel et al. 2012). Kreidberg et al. (2012) observed the upper limit of the "m-gap" $M_{\text{m-gap}} = 2\text{–}5\ M_\odot$ in the distribution, and the most reliable value for the lowest BH mass is above $M_{BH} > 6\ M_\odot$, with a peak at $\approx 2.1\ M_\odot$ for 16-1 X-ray binaries.

However, at the end of this section, based on the available data, it is useful to present the statistical distribution of the masses of some important and credible compact objects that fall into the "m-gap" region (Fig. 4).

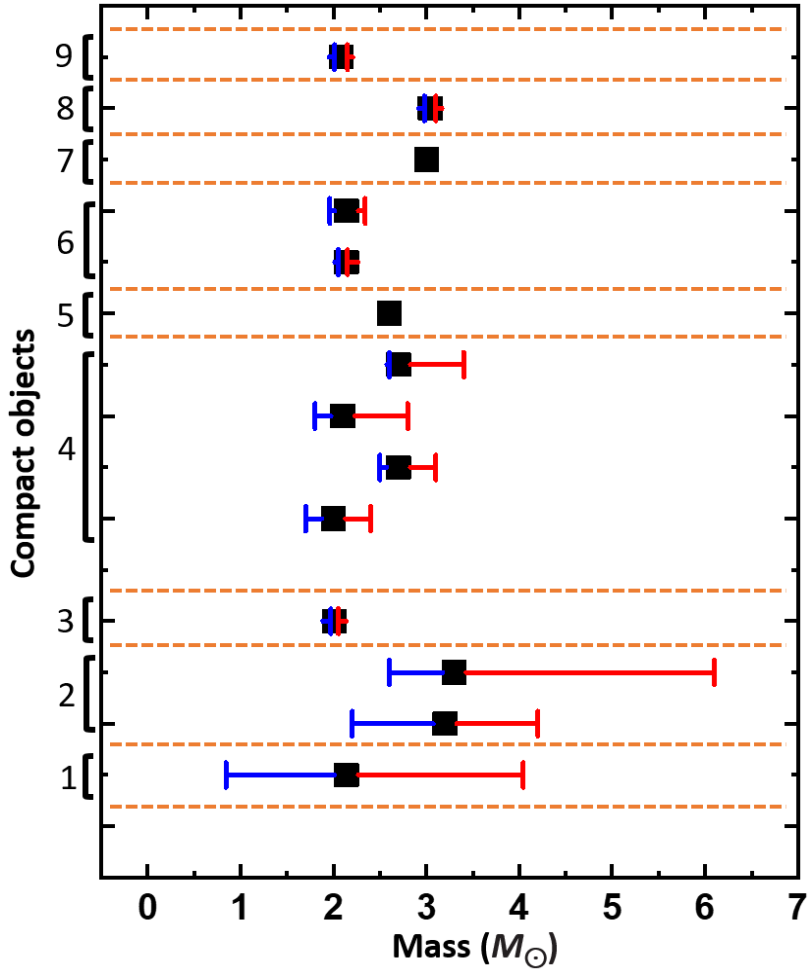

**Fig. 4.** Compact Steller masses filling the "m-gap": 1 – OGLE3-ULENS microlensing events (PAR-03) (Wyrzykowski and Mandel 2020); 2 – J05215658 (Thompson et al. 2019); 3 – J0348+0432 (Antoniadis et al. 2013); 4 – GW 190814 (Galaudage et al. 2021); 5 – GW190814 (Fattoyev et al.2020); 6 – J0740+6620 (68.40 % credibility) (Cromartie et al. 2020); 6 - J0740+6620 (95.4 % credibility) (Cromartie et al. 2020), 7 – 4U 1545-47 (Orosz et al. 1998); 8 – V723 Mon (Jayasinghe et al. 2021); 9 – J0740+6620 (Fonseca et al. 2021).

It is worth mentioning that the "m-gap" range between the highest NSs and lowest BHs masses is inconsistent, and this is summarized in Table 1. Different groups considered this region with different mass values. The statistical summary shows that $M_{\text{m-gap}} = 2\text{–}5\ M_\odot$. is considered to be the most reliable "m-gap" value.

Table 1. Summary of different values of the "m-gap" region between the highest NS masses and lowest BH masses considered by different groups.

|   | **Mass gap range** | **References** |
| --- | --- | --- |
| 1 | 2–5 $M_\odot$ | Özel et al. (2010), Özel et al. (2012), Belczynski et al. (2012), Kreidberg et al. (2012), Abbott et al. (2019), Abbott et al. (2020c), Yang et al. (2020), Abbott et al. (2021b), Burrows and Vartanyan (2021), Zevin et al. (2020) |
| 2 | 2–6 $M_\odot$ | Heida et al. (2017) |
| 3 | 3–5 $M_\odot$ | Gelino and Harrison (2003) |

In Fig. 4, the errors represented by blue and red line indicate the lower and upper possible mass limits for compact objects, respectively, in relation to the peak mass value. Statistical analysis showed maximum number of compact objects falling in the mass region 2.1-2.7 $M_\odot$, which were usually considered to be supermassive NSs. However, with greater credibility, few of the compact objects that reach this limit are considered to be BHs with a lower mass. Unfortunately, the EoSs and microscopic properties of these objects are not well understood.

## 4. GRAVIDYNAMICS OF DENSE COLLAPSARS

In gravidynamics, new properties of compact gravitating collapsars of stellar mass may arise if the energy densities of the gravitational field become close to or exceed the nuclear density value $\rho_{\text{nucl}} = 2.8 \times 10^{14}$ g cm$^{-3}$. The fundamental interaction between particles are dominated by the gravitational filed if the density of matter is much higher than $\rho_{\text{nucl}}$. The phase transition of matter to quark-gluon plasma during formation of dense objects can be directly associated with such a dynamical gravitational description. The strong field of collapsars which is an analogue of the BH in GR is investigated in such totally non-metric, dynamical model of gravitational interaction theory. In the case of extremely strong collapsar field in gravidynamics, a region filled by matter ("a bag") must have a radius equal to $r^* = GM_Q c^2 \lesssim 10$ km at the total collapsar mass $M_Q \lesssim 7\ M_\odot$ (Sokolov 1992, 2019).

The strong static field of a collapsar in gravidynamics provides specific estimation of mass 6.7 $M_\odot$ (Sokolov 2019; Sokolov and Zharykov 1993), which describes the quark-gluon plasma in the "bag". The color force in quantum chromodynamics (QCD) describes the self-bound state of the quark-gluon plasma. The "macroscopic" constant of the color forces consists of 3 constants of the electromagnetic interaction ($\alpha_{\text{QED}} \approx 0.0073$) at a distance of $r \sim 1\ fm$ from the center of such a "bag" with the radius $\lesssim 10$ km. In such a state, the density $\varepsilon(r)/c^2$ approaches a value of the order of $\rho = 5.4 \times 10^{52}$ g cm$^{-3}$, and the total energy (mass) in the confined small sphere of $r = 1\ fm$ is equals to $7 \times 10^{14}$ g cm$^{-3}$. The total mass of $M_Q \approx 6.7\ M_\odot$ was calculated based on the metric and dynamic gravitation theories for extremely compact object in gravidynamics (Sokolov 2019; Sokolov and Zharykov 1993), using the expression $M_Q = 6.64 M_\odot \left(\dfrac{2\rho_{nucl}}{4B/c^2}\right)^{1/2}$, where the surface consists of a strange self-bound matter of a "bag" of radius $r^* = GM_Q c^2 \lesssim 10$ km. The results in mass distributions are highlighted with a sharp red line at $M_Q \approx 6.7\ M_\odot$ (Fig. 1a). This selected mass value of 6.7 $M_\odot$ was obtained once considering the particular value of the "bag" constant $B = 79.925$ $MeV\,fm^{-3}$ for quark-gluon plasma "bag" model in QCD. The limiting equation of state $P_Q = 1/3(\varepsilon - 4B)$ for the quark configurations was considered, where $\varepsilon$ is the total energy density inside the massive quark-gluon "bag" with the radius $r^* = GM_Q c^2 \lesssim 10$ km. Such a concept was discussed by Witten (1984) and revised by Sokolov and Zharykov (1993). The limitation $2.3\,\rho_{nucl} \gtrsim \rho_{QGP}(P_Q = 0) \gtrsim 1.7\,\rho_{nucl}$ fixes the mass and radius of a cold quark star in gravidynamics within the limits $6.21\ M_\odot \lesssim M_Q \lesssim 7.25\ M_\odot$ and $9.16$ km $\lesssim R_{GQP} \lesssim 10.69$ km, respectively. The lowest value of the total mass and QGP bag radius follow from the condition that at the density $\varepsilon(r) = (4B/c^2) R^2_{QGP}\, r^{-2}$ on the bag surface is equal to $\varepsilon/c^2 \approx 2.3\rho_{nucl}$. The surface consists of the strange self-connected (i.e., stable at $P_Q = 0$) matter. As $\varepsilon$ increases towards the center of the bag, consequently, all other kinds of quarks become defrozen. Then in gravidynamics, a cold quark star with the strange surface is considered, if the density on the surface does not exceed $\approx 2.3\rho_{nucl}$.

The model of gravidynamics is based on the concept of gravitational interaction of the matter field which is defined by the energy. The gravitating mass of a compact object is described by the gravitational field energy. The concept is similar to electromagnetic mass of electron and the electron-electron interaction defined by electrodynamics. The force is mainly specified only by the tensor part of the gravitational field in the relativistically weak field mode, when $r \gg r^* = GM_Q c^2$ (Sokolov 2019; Sokolov and Zharykov 1993). However, if the energy density of the field itself approaches the nuclear density $\rho_{nucl}$ (for $r \approx r^*$) of a dense compact object with the strong gravitational field, the role of the scalar component of the field, which is characterized as repulsion, increases. In this particular condition, the physical properties of the total mass $M_Q$ of such an extremely dense compact object are considered as a quark star in gravidynamics. The half of the quark star is composed by the field energy of scalar-tensor components around the "bag" with radius $r^*$. This suggests the basic observational consequence confirming the unification of gravidynamics and QCD for the existence of a selected collapsar mass with a value of $M_Q \approx 6.7\ M_\odot$.

In sufficiently weak fields (for $r \gg r^*$), the relativistic effects in gravidynamics and GR acts in similar way. GR describes the gravitation as a tensor field with the massless spin 2 graviton particle in flat space-time. In such a theory, all particles that carry energy interact with the graviton. In quantum field theory, the potential is usually determined by the fact that two particles exchange gravitons. However, in the strong gravitational potential, a consistent dynamic description of the field (gravidynamics) can give a distinct result. This is possible only when the size of a dense compact object becomes close to $GM/c^2$. This is a possible state of field when the energy density of the field itself becomes comparable with the energy density of matter. In gravidynamics, the observational properties and a stable quark star with strong (for $r \sim r^*$) field are nevertheless determined by the scalar component of the field of "gravitons" with spin 0.

The direct observational consequences of such a consistent dynamical description of gravitation with two field components of spin 0 and 2 is a matter of confirmation. This strongly suggests an explanation for the possibility of the second peak reported by Özel et al. (2012) in the mass distribution of compact stellar objects (Fig. 1a). In gravidynamics, contribution of scalar emission of gravitational energy is accurately considered. The experimental evidence of such an effect was attempted by Sokolov (1992) by considering PSR 1913+16. The discrepancy of secular period changes $dP/dt$ in the binaries from GR was also observed. In particular, the observed $dP/dt$ by Weisberg et al. (2010) is higher, and it indicates unaccounted contribution, which should be taken into account.

The relationship of CCSNs and GRBs is needed for the understanding of the existing "m-gap" between $M_{\text{m-gap}} = 2$–$5$ $M_\odot$. The specific CCSN explosion mechanism itself is critical for understanding the "m-gap" in the mass distribution function. At critically high gravitational potential, the part of the stellar matter does not receive enough energy to escape the gravitational potential well of a newly formed NS, and it falls back onto the core. This is the fallback, which can be also connected to long-duration GRBs. After the formation of a BH, the gravitational potential well becomes infinitely deep. In this case, there would not be sufficient finite energy to leave matter, consisting of elementary particles, against an infinitely deep gravitational potential, and this condition prohibits the explosion of a CCSN.

A model of core collapse massive star progenitors through the core bounce was proposed by Janka (2012), in which the responsible candidate is governed by neutrino transport mechanism. In this model, first, an NS is formed, and further the "fallback" leads to BH formation, generating a long-duration GRB and neutrino emission. The calculations indicated that such an "m-gap" can indeed provide constraints on the physical mechanism of CCSN explosion. The observed optical companion masses are found mostly in the mass range 0.1–1 $M_\odot$, but with a peak at 0.6 $M_\odot$. Under such conditions, none of the available common envelope models allow for the formation of the observed population of Galactic BH transients with masses highlighted by Özel et al. (2012) (Fig. 1). However, the small masses of optical companion ($M_{\text{opt}} = 0.6$ $M_\odot$) is difficult to explain. Belczynski et al. (2012) showed that stars with a mass above $M > 2$ $M_\odot$ are the most likely companions of Galactic BHs. However, the model showed that stars with the mass $M = 1$ $M_\odot$ are the most probable companions of BHs.

In the NS and BH mass distribution, the characteristic absence of compact objects exists in the mass range $M_{\text{m-gap}} = 2$–$5$ $M_\odot$. This is close to the interval between the maximum mass of NSs and minimum mass of BH candidates. In this case, the interval between the mass distribution of two peaks are close to 1.4–6.7 $M_\odot$. This suggests that the "m-gap" begins indeed with a sharp decline of mass of compact objects both from the NSs ($M_{\text{NS}} = 1.4$ $M_\odot$), and the BH candidates side ($M_{\text{BH}} = 6.7$ $M_\odot$), which is shown in Fig. 1a. A model proposed by Belczynski et al. (2012) described the existence of lowest mass of BHs above $M_{\text{BH}} > 5$ $M_\odot$ in correspondence with the fact that the maximal mass of NSs is less than $M_{\text{NS}} < 2$ $M_\odot$. A model proposed BH masses in the range of $M_{\text{BH}} = 5$–$15$ $M_\odot$. This range supports the existence of the "m-gap", which is consistent with masses of BHs in the Galactic binaries (Belczynski et al. 2012). This model additionally describes that BHs may exist in this "m-gap" due to the CCSN explosion. This model indeed supports the existence of BHs mass around $M_{NS} \approx 3\ M_\odot$. However, this finding is inconsistent with the observations, since it does not fit well with the model, which indicate that BH masses would be shifted to a very narrow range with a peak at $M_{BH} \approx 3\ M_\odot$. This result also contradicts the model proposed by Özel et al. (2012) (Fig. 1), in which the mass of BHs turns out to be above $M_{BH} \approx 5\ M_\odot$. It is also important to consider that limit mass of observed NSs is up to $M_{NS} \approx 2\ M_\odot$, with restricted EoS, $P = \varepsilon$ (Shapiro and Teukolsky 1983). Therefore, the prediction based on the theory of strong interaction failed to describe NSs and BHs with the mass 3.2 $M_\odot$.

## 4.1. DISCUSSION

The mass of the supergiant optical star HDE 226868 (Cygnus X-1) with a relativistic companion was studied and determined by Kopylov and Sokolov (1984) and Sokolov (2019). It turned out that with the total mass $M_{opt} = 19.5\ M_\odot$, the mass of the BH candidate must be $M_{BH} \gtrsim 6.5\ M_\odot$. This lower mass $M_x = 6.5\ M_\odot$ limit (Sokolov 1987) of the relativistic object in Cygnus X-1 is close to the low mass cut-off ($M_c$) at $M_c = 6.32\ M_\odot$ in the exponential mass distribution with the peak value of $M_{peak} = 6.7\ M_\odot$ (Özel et al. 2010). The properties of this relativistic compact object are considered as BH with singularity and event horizon. However, it may be possible that dense compact object with masses $M \approx 2$–$5\ M_\odot$ have some unusual properties such as size, surface and EoSs. The lack of the pulsar in the objects of mass $M = 6.7\ M_\odot$ in the mass distribution does not confirm the surfaceless state of this object. This indicates that the objects are not pulsating. The study of the surface of this object and the verification of GR predictions in the strong-field regime is compulsory to confirm the properties of this object (Maselli et al. 2015; Psaltis 2008).

The physical properties of collapsars could be explained by emission anisotropy of GRBs and black body spectra of GRBs. Linearly polarized emission of GRBs can, in principle, be explained by the direct appearance of strong magnetic field of a collapsar, resulting from a CCSN explosion (Burrows and Vartanyan 2021). It may be associated to radiation transfer in a medium with high magnetic field of value $\sim 10^{12} - 10^{16}$ Gauss from the surface of the compact object (Mao and Wang 2013). Moreover, a cyclotron energy, $E_{cyc} = 21.7\ (+1.9/-1.6)$ keV with gravitational red-shift $z_{grav} = 0.318$ for GRB 011211 might be explained by the direct manifestation of the surface magnetic field of $\sim 10^{12}$ Gauss in the GRB gamma-ray photon spectrum with $z_{GRB} = 2.140$ (Frontera et al. 2004).

The problem of massive NSs is also actively reported and discussed in the above sections. However, the EoSs of the massive star is not deterministic and the core composition of such an object is not known. It is suggested that the matter core of the NS exists in the ultra-dense state (Weber et al. 2014). In this region, the existence of quark deconfinement and stable strange quark matter becomes feasible. An allowed maximum mass of the NSs is also not well understood from the observational point of view. The description of EoSs becomes complicated for these objects, which showed the first peak in the mass distribution (Özel et al. 2012). Such a complication further indicates doubtful existence of NS in the mass distribution (Weber et al. 2014). Thus, the peak of 1.48 $M_\odot$ in the mass distribution is also a matter of debate and investigation. The masses of NSs of $\sim 2\ M_\odot$ are permissible in this mass distribution. However, such compact objects are heavy to consider them as NSs (Lattimer 2012). This further, hints that such compact objects are neither NSs nor BHs. We propose to connect the CCSN explosion and GRBs measurements in the near future in order to confirm the properties of such exotic compact objects, the peak of which is located at 1.48 and 6.7 $M_\odot$ in the mass distribution.

## 5. SUMMERY

The distribution of gravitational mass of neutron stars and black holes (collapsars) with the formation of the "mass gap", $M_{\text{m-gap}}$ = 2–5 $M_\odot$ between these collapsars was analyzed. In this distribution, the maximum likehood of neutron stars and black holes was located with the masses of $M_{\text{NS}}$ = 1.4 $M_\odot$ and $M_{\text{BH}}$ = 6.7 $M_\odot$, respectively. However, recent observational results have indicated the filling of the "mass gap". The notion of gravidynamics was proposed to settle the peak mass at 6.7 $M_\odot$ and "mass gap" problem in the mass distribution. This concept was based on the non-metric scalar-tensor model of gravitational interaction with the consideration of localizable field energy. The gravidynamics model showed that the compact relativistic object of $M$ = 6.7 $M_\odot$ was constituted by matter of quark-gluon plasma of radius $r^* = GM_Q/c^2 \approx 10$ km. It was conjectured that the total measurable gravitational mass of such a dense compact object consists of both matter and field, which is a combination of scalar-tensor components. It was suggested that the polarized emission of long gamma-ray bursts and their response in the form of a blackbody component in the spectra could be explained by physical properties of collapsars.

## ACKNOWLEDGMENTS
The authors are grateful to T. N. Sokolova for her help in preparing the manuscript.
## FUNDING
No funding was received for conducting this study.
## CONFLICT OF INTERES
The authors declare that there is no conflict of interest regarding the publication of this article.

The work was performed as part of the government contract of the SAO RAS approved by the Ministry of Science and Higher Education of the Russian Federation.